\documentclass[%
 reprint,
superscriptaddress,
 amsmath,amssymb,
prb,
]{revtex4-2}

\usepackage{graphicx}
\usepackage{dcolumn}
\usepackage{bm}
\usepackage{multirow}
\usepackage{tikz}
\newcommand{\tikzxmark}{%
\tikz[scale=0.2] {
    \draw[line width=0.7,line cap=round] (0,0) to [bend left=6] (1,1);
    \draw[line width=0.7,line cap=round] (0.2,0.95) to [bend right=3] (0.8,0.05);
}}
\usepackage{chngcntr}

\begin{document}

\preprint{APS/123-QED}

\title{First principles study of [111]-oriented epitaxially strained\\Rare-Earth Nickelate NdNiO$_3$}

\author{Alexander Lione}
\affiliation{%
 Condensed Matter Physics, Physics Department, Durham University, Durham, United Kingdom
}%


\author{Jorge Í\~niguez-González}
\affiliation{%
 Smart Materials Unit, Luxembourg Institute of Science and Technology, Avenue Des Hauts-Forneaux 5, L-4362 Esch/Alzette Luxembourg
}%
\affiliation{%
 Department of Physics and Materials Science, University of Luxembourg, Rue du Brill 41, L-4422 Belvaux, Luxembourg
}%
\author{Nicholas C. Bristowe}
\affiliation{%
 Condensed Matter Physics, Physics Department, Durham University, Durham, United Kingdom
}%

\date{\today}

\begin{abstract}
Density functional theory is used to investigate the effect of biaxial strain on the structural, electronic and magnetic properties of [111]-oriented NdNiO$_3$, as a representative of the rare-earth perovskites that undergo metal-to-insulator transitions. We find that this constraint on the system induces unique structural phase transitions not previously observed under the well-studied bulk or [001]-oriented strained systems. We also report unique electronic behaviour, including amplification of the electronic band-gap with tensile strain, and insulating, charge-ordered phases with non-orthorhombic tilt patterns. To provide clarity to the trends we observe, we also investigate the coupling between the breathing mode and strain, where we observe certain strains to directly favour and disfavour the creation of the breathing mode (and thus the associated charge-ordering). The amplification of the band gap with strain is understood in terms of a cooperative coupling between the elastic constraint and octahedral breathing, which expands on the previously reported triggered mechanism mediated by octahedral tilting.
\end{abstract}

\maketitle


\section{\label{sec:level1}Introduction}
The rare-earth nickelates $R$NiO$_3$ (where $R$ denotes a rare-earth element) are a fascinating playground for novel structural and electronic phase transitions. With exception to La \cite{LaNiO3_Expt}, all rare-earth nickelates undergo a metal-insulator transition (MIT) with decreasing temperature \cite{Nickelate_Expt_MIT_1992,MIT_general}. Concomitantly, the symmetry lowers from an orthorhombic $Pbnm$ phase to a monoclinic $P2/c$ phase \cite{Nickelate_Expt_Breathing_1992,Nickelate_Expt_Breathing_2000,Nickelate_Expt_Breathing_1999} with charge ordering on the Ni sites, which change from an oxidation state of $2\text{Ni}^{3+}\rightarrow\text{Ni}^{+3+\delta}+\text{Ni}^{+3-\delta}$ \cite{Nickelate_Expt_2009}. Research into the unique properties of the rare-earth nickelates has led to a deeper understanding of several fundamental concepts, including the nature of poorly conducting materials \cite{bad_conductivity}, multiferroicity \cite{NNO_mag_order_expt_1}, MIT triggering via structural control \cite{Triggered_MIT}, and MIT tuning via elastic fluctuations \cite{elastic_MIT}. Such phenomena have also been directly applied to designing novel electronic devices with unique properties, including artificial neuromorphic systems utilising conductance switching via Ni oxidation and reduction \cite{synaptic_transistor}, artificial neural networks built from nickelate memory capacitors (driven by rapid changes in resistivity at the MIT) \cite{nickelate_AI}, and sensing devices reliant on Ni charge transfer \cite{nickelate_sensors}.%

The strain enforced on an epitaxial film provides a strategy to manipulate the properties of nickelates. [001]-oriented epitaxial strain allows for direct control of the MIT temperature, which has been shown to decline with increasing compressive \cite{Strained_SNO_film} and tensile strain \cite{Strain_NNO_2002,Strain_SNO_2000}. Density functional theory (DFT) studies have predicted that rare-earth nickelates undergo multiple phase transitions, including an insulator-metal transition (IMT) under compressive strain, and an IMT followed by a second MIT (via orbital ordering) under increasing tensile strain \cite{Nickelate_001_Strain_DFT}. 

[111]-oriented perovskites remain significantly less studied. Straining the [111]-oriented crystal applies a unique constraint to the cubic perovskite cell, breaking the symmetry of the four equivalent [111]-directions into three symmetrically-equivalent (if assuming isotropic in-plane strain) and one symmetrically-inequivalent [111] axis. In other perovskites, [111]-strain has led to vastly different strain-phase diagrams in comparison to [001]-strain \cite{111_General_orthorhombic_perovskite,111_General_perovskite}. In the case of the rare-earth nickelates, experimental studies have observed unique MIT variation under tensile [111]-strain \cite{111_NNO_expt,NNO_111_2020}, the creation of $A$-site driven `polar metals' \cite{Polar_metals}, and novel orbitally-ordered phases in layered superlattices \cite{Mott_NNO_111}. It is challenging to identify the underlying cause of these observations, whether it be strain, mode matching, the polar catastrophe, quantum confinement, or magnetic frustration, all of which may be present in [111]-nickelate systems.

This study aims to provide clarity to [111]-oriented rare-earth nickelate systems by isolating the effect of [111]-strain. We report a first-principles investigation into the effect of [111]-strain on the rare-earth nickelate NdNiO$_3$ (NNO), whose qualitative properties are representative of the rest of the rare-earth nickelates (except La). We present a description of the essential phases and properties of NNO in bulk, followed by a detailed report of its structure and electronic properties as a function of [111]-strain. We demonstrate that both compressive and tensile strain generate previously-unseen nickelate phases with unique features. These compliment our previous understanding of how various structural distortions are coupled to one-another, as well as how they influence electronic properties such as the band gap. To obtain a deeper understanding of how the nickelates respond to epitaxial strain, we also investigate precisely its effect on the hallmark structural feature of the nickelates--- the breathing mode. We measure the energy penalty of forming the breathing mode at a range of different strains in both the [111]- and [001]-basis to understand which are cooperative or competitive with its occurrence. Furthermore, we search for answers regarding the driving force behind the polar nickelate phases seen in [111]-oriented interface systems \cite{Polar_metals}. To do this, we test our systems for ferroelectricity, by checking for inversion symmetry breaking and trends in the dielectric response with strain.

\section{\label{sec:level1}Methods}

\begin{figure}
    \centering
    \includegraphics[width=1\linewidth]{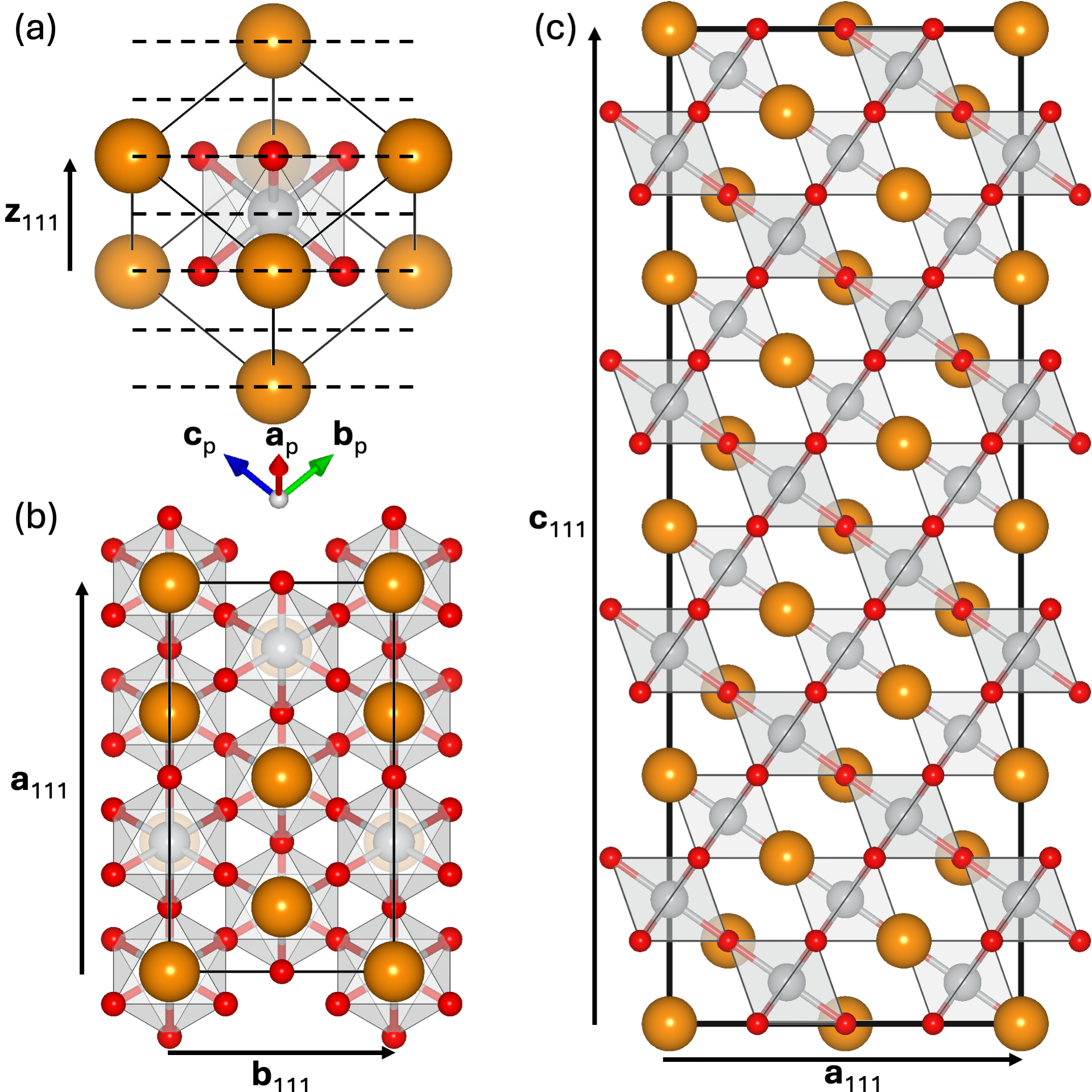}
    \caption{Visualisation of [111]-oriented perovskites. (a) Rotated $Pm\bar{3}m$ primitive perovskite cell such that the $\textbf{z}_{111}$ ($\hat{\textbf{c}}_{111}$) direction lies vertical. Horizontal dashed lines indicate (111) layers. (b), (c) 120 atom [111] NdNiO$_3$ unit cell, viewed from the $\textbf{a}_{111}-\textbf{b}_{111}$ plane (in which bi-axial strain is applied), and $\textbf{a}_{111}-\textbf{c}_{111}$ plane respectively.}
    \label{fig:111_cell_visual}
\end{figure}

We performed density functional theory (DFT) calculations within the generalised gradient approximation (GGA), utilising the PBEsol exchange-correlation functional \cite{PBEsol}, and projector augmented wave (PAW) generated pseudopotentials \cite{PAW} within the PBE scheme \cite{PBE}, implemented in the Vienna Ab-initio Simulation Package (VASP) \cite{VASP_1,VASP_2,VASP_3}. The Nd $5s$, $5p$, $5d$ and $6s$; Ni $3p$, $3d$ and $4s$; and O $2s$, $2p$ electrons were explicitly included as valence, while all other electrons (including Nd's $4f$, assuming a 3+ ionisation state) were frozen in the ionic cores. For ionic relaxations, the plane-wave energy cut-off was set to 900 eV, while the K-point grid was fixed to a $3\times 5 \times 1$ $\Gamma$-centered mesh for the 120 atom unit cell in Fig. \ref{fig:111_cell_visual}(b), (c). For calculations of high-resolution band structures and the dielectric tensor, the plane wave energy cut-off was set to 550 eV in order to make them computationally tractable. The dielectric tensor was calculated via density functional perturbation theory \cite{VASP_DE}. 
A Hubbard-$U$ correction \cite{HubbardU} of 2 eV was applied to the Ni-$3d$ orbitals, as suggested by previous studies to obtain the correct bulk Nickelate ground state \cite{Complete_Phase_Diagram_Nickelates,Nickelate_U_Calc}. To quantify structural distortions we used symmetry adapted mode analysis, as implemented in the ISOTROPY package \cite{ISOTROPY}.

The [111]-oriented nickelate unit cell is visualised in Fig. \ref{fig:111_cell_visual}(a)---(c). Primitive pseudocubic perovskite directions are denoted by $\hat{\textbf{a}}_{\text{p}}$, $\hat{\textbf{b}}_{\text{p}}$ and $\hat{\textbf{c}}_{\text{p}}$ in Fig. \ref{fig:111_cell_visual}(a), while the lattice vectors of the [111]-oriented unit cell are denoted by $\textbf{a}_{111}$, $\textbf{b}_{111}$ and $\textbf{c}_{111}$ in Fig. \ref{fig:111_cell_visual}(b), (c). The lattice vectors may be expressed in terms of primitive cubic cell vectors: $\textbf{a}_{111} = \textbf{a}_{\text{p}}+\textbf{b}_{\text{p}}-2\textbf{c}_{\text{p}}$, $\textbf{b}_{111} = -\textbf{a}_{\text{p}}+\textbf{b}_{\text{p}}$, $\textbf{c}_{111} = 4\textbf{a}_{\text{p}}+4\textbf{b}_{\text{p}}+4\textbf{c}_{\text{p}}$ (Fig. \ref{fig:111_cell_visual}(b), (c)). Note that this large supercell is necessary for expressing both the NdNiO$_3$ structure and complex magnetic order. When $\textbf{a}_{111}, \textbf{b}_{111}$ have the same applied strain, all three primitive directions experience an equivalent geometric constraint.

\begin{figure}
    \centering
    \includegraphics[width=1\linewidth]{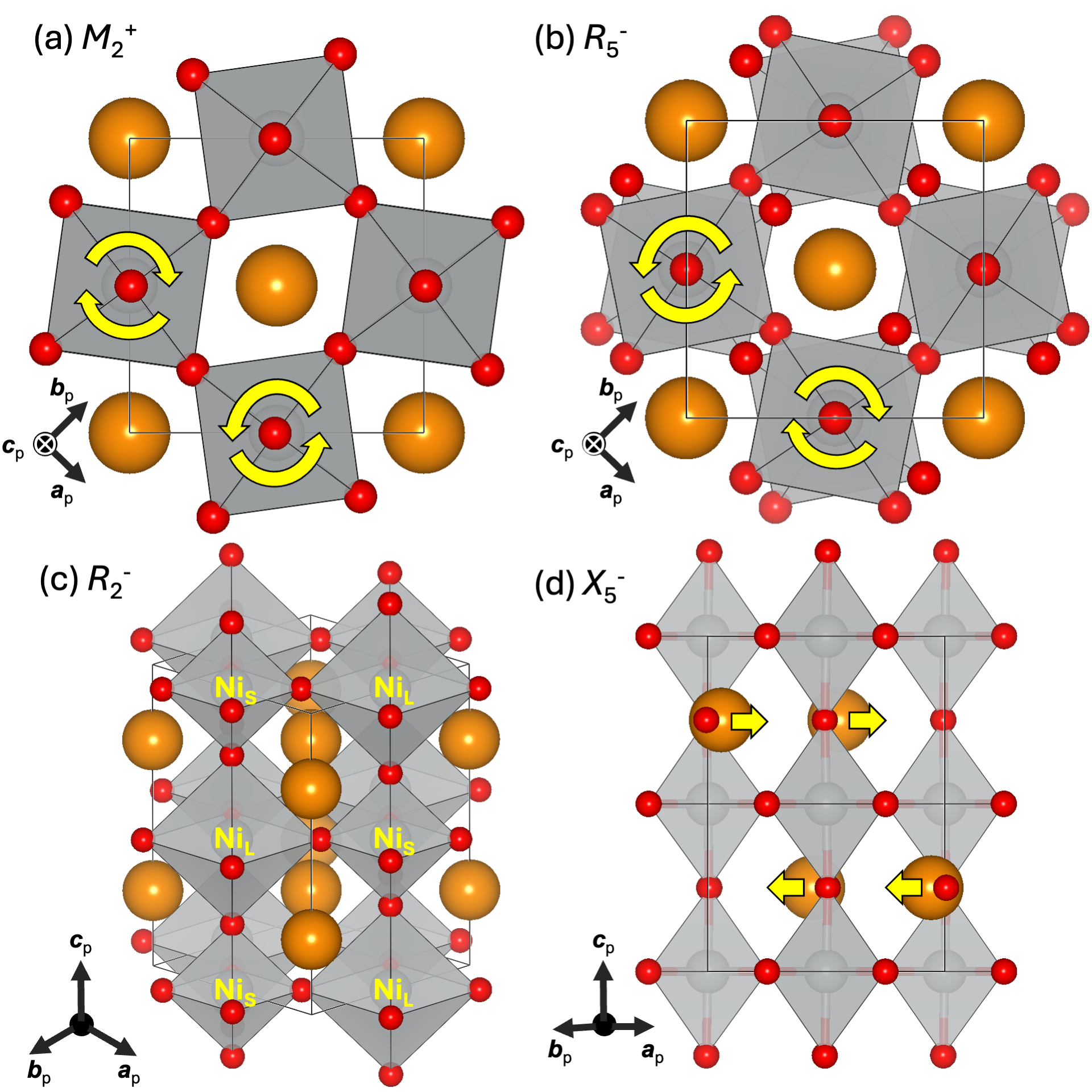}
    \caption{Primary structural distortions in $R$NiO$_3$, visualised in the [001]-basis, denoted by corresponding irrep labels and order parameter directions (with respect to the $A$-site origin perovskite cell): (a) in-phase octahedral tilt ($M_2^+(0,0,c)$) (viewed from $\hat{\textbf{c}}_{\text{p}}$), (b) anti-phase octahedral tilt ($R_5^-(0,0,c)$) (viewed from $\hat{\textbf{c}}_{\text{p}}$), (c) breathing distortion ($R_2^-(a)$), (d) antipolar $A$-site motion ($X_5^-(0,0;0,0;a,a)$). Gold spheres indicate Nd sites, grey spheres indicate Ni sites, and red spheres indicate O sites. 
    }
    \label{fig:modes_visual}
\end{figure}

In this study, we simulated four different structural phases. We tested both charge-ordered and non-charge-ordered inputs for two types of octahedral tilt order (written in [001]-basis Glazer notation for clarity): $a^-a^-c^+$ (corresponding to charge-ordered, monoclinic $P2/c$ and non-charge-ordered, orthorhombic $Pbnm$ phases), and $a^-a^-a^-$ (corresponding to the charge-ordered, triclinic $P\bar{1}$ phase, where the appearance of the breathing mode breaks the symmetry to $a^-b^-c^-$, and the non-charge-ordered, rhombohedral $R\bar{3}c$ phase). A summary of these phases and the corresponding symmetry reduction induced by [111]-strain is presented in Table \ref{tab:space_groups_summary}, and the most significant structural distortions with their corresponding irrep (irreducible-representation) label in Fig. \ref{fig:modes_visual}(a)---(d). 
We also note that these tilt patterns often reverted to higher symmetry patterns after structural relaxation.

\newcommand{\xmark}{\ding{55}}
\begin{table}[]
\caption{Space groups of various nickelate phases, where tilt patterns are denoted in Glazer notation, charge ordering corresponds to the presence of irrep $R_2^-$, and strain irrep labels correspond to the presence of [001]- ($\Gamma_3^+(a,0)$) and [111]-strain ($\Gamma_5^+(a,a,a)$). $^*$Broken symmetry allows breathing. $\dagger$Broken symmetry allows for $a^-a^-c^-$ tilt pattern. $^{\ddagger}$Broken symmetry allows for $a^-b^-c^-$ tilt pattern.}
\label{tab:space_groups_summary}
\begin{ruledtabular}
\begin{tabular}{cccc}
Tilt pattern & Charge Ordering & Strain & Space group \\ \hline
\multirow{6}{*}{$a^-a^-c^0$} &  &  $\tikzxmark$& $Imma$ \\
 &  $\tikzxmark$& $\Gamma_3^+$ & $Imma$ \\
 &  & $\Gamma_5^+$ & $C2/m^*$ \\ \cline{2-4}
 &  &  $\tikzxmark$& $C2/m$ \\
 & $\checkmark$ & $\Gamma_3^+$ & $C2/m$ \\
 &  & $\Gamma_5^+$ & $C2/m$ \\ \hline
\multirow{6}{*}{$a^-a^-c^+$} &  &  $\tikzxmark$& $Pbnm$ \\
 &  $\tikzxmark$& $\Gamma_3^+$ & $Pbnm$ \\
 &  & $\Gamma_5^+$ & $P2/c^*$ \\ \cline{2-4}
 &  &  $\tikzxmark$& $P2/c$ \\
 & $\checkmark$ & $\Gamma_3^+$ & $P2/c$ \\
 &  & $\Gamma_5^+$ & $P2/c$ \\ \hline
\multirow{6}{*}{$a^-a^-a^-$} &  &  $\tikzxmark$& $R\bar{3}c$ \\
 &  $\tikzxmark$& $\Gamma_3^+$ & $C2/c^{\dagger}$ \\
 &  & $\Gamma_5^+$ & $C2/c^{\dagger}$ \\ \cline{2-4}
 &  &  $\tikzxmark$& $P\bar{1}^{\ddagger}$ \\
 & $\checkmark$ & $\Gamma_3^+$ & $P\bar{1}^{\ddagger}$ \\
 &  & $\Gamma_5^+$ & $P\bar{1}^{\ddagger}$
\end{tabular}
\end{ruledtabular}
\end{table}

Experimental studies have shown that NdNiO$_3$ has a particularly complex magnetic order, with non-collinear AFM orders on both Ni and Nd sites at ground state \cite{NNO_mag_order_expt_1,NNO_mag_order_expt_2}. For simplicity, we choose to omit the Nd magnetic moment entirely (hence the omission of $4f$ electrons), as it only exhibits long range order below 30 K \cite{Nickelate_Expt_magnet_1994,Nickelate_review_1997,Strained_SNO_film} and is not essential for the MIT and Ni charge ordering which occur at around 200 K \cite{Nickelate_Expt_MIT_1992, Nickelate_Expt_Breathing_1992,NNO_mag_order_expt_1,Nd_mag_order_1}. For computational tractability we choose not to simulate non-collinear orders on the Ni sites, as in previous bulk studies, collinear $T,S$-AFM $\uparrow \uparrow \downarrow \downarrow$ spin-chains have been demonstrated to be lower in energy than non-collinear orders \cite{Complete_Phase_Diagram_Nickelates}, and sufficiently recreate the experimental properties. With the appearance of Ni charge ordering, $T,S$-AFM spin chains produce 0 moments on the higher valence Ni$^{+3+\delta}$ sites (sites with moments correspond to lower valence Ni$^{+3-\delta}$). This leads to alternating magnetic and non-magnetic Ni layers along the [111]-direction, where $T$-AFM consists of FM in-plane layers, while $S$-AFM consists of AFM in-plane layers. In this study, we tested these two magnetic orders on insulating phases, whereas we modeled the metallic phase with FM ordering as is commonly done in the DFT literature \cite{Nickelate_001_Strain_DFT,Complete_Phase_Diagram_Nickelates,Triggered_MIT}.

\begin{figure}
    \centering
    \includegraphics[width=1\linewidth]{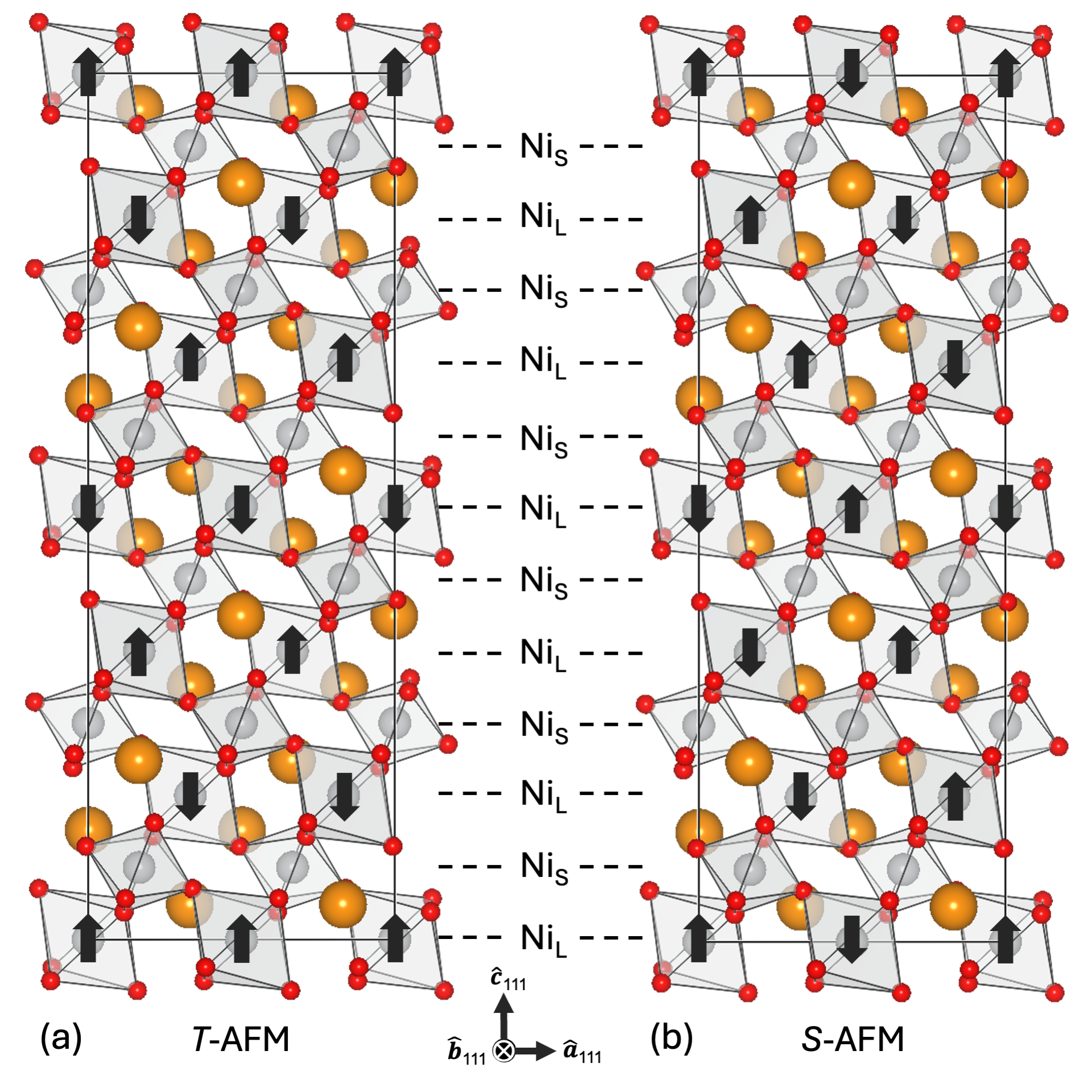}
    \caption{Magnetic orders of insulating, charge-ordered phases. Small and large Ni valence sites are denoted by Ni$_{\text{S}}$ (Ni$^{+3+\delta}$) and Ni$_{\text{L}}$ (Ni$^{+3-\delta}$) respectively. Ni$_{\text{S}}$ sites have no magnetic moment. (a) $T$-AFM, (b) $S$-AFM magnetic orders.}
    \label{fig:Magnetic_visual}
\end{figure}

The $T$- and $S$-AFM magnetic orders in this basis are depicted in Figs. \ref{fig:Magnetic_visual}(a) and \ref{fig:Magnetic_visual}(b) respectively. To apply [111]-strain to this system, we fixed the in-plane lattice parameters $\textbf{a}_{111},\textbf{b}_{111}$ to a value by fixing the corresponding components of the stress tensor, using a stress constraining geometry optimisation patch in VASP \cite{VASP_Strain_Patch}. Both in-plane lattice vectors were set to the same (pseudocubic) lattice parameter, and were varied from a highly compressive 3.6$\:$Å to a highly tensile 4.0$\:$Å regime.

\section{\label{sec:level1}Results and Discussion}

\subsection{Bulk NdNiO$_3$}

In bulk, we find the $P2/c$ structure to be the ground state, as seen in previous DFT studies using similar parameters \cite{Complete_Phase_Diagram_Nickelates,Nickelate_001_Strain_DFT}. We find the $T$-AFM magnetic order to be marginally lower in energy than the $S$-AFM order in our calculations. The metallic $Pbnm$-FM phase is higher in energy, while the $a^-a^-a^-$ phases are higher in energy still. The respective energies of each bulk phase are reported in Table \ref{tab:Bulk_Energies}. The band gap of the ground state $P2/c$-$T$ phase was found to be 0.47 eV, comparable to experiment \cite{NNO_BG} and previous DFT studies \cite{Complete_Phase_Diagram_Nickelates}. Mode magnitudes and lattice parameters (see Appendix A for details) were also found to be comparable to low temperature synchrotron data \cite{Nickelate_Expt_2009}. The ground state [111]-basis lattice parameters (with pseudocubic equivalent included in brackets) were found to be $|\textbf{a}_{111}|=9.26$ Å ($/\sqrt{6} = 3.78$ Å), $|\textbf{b}_{111}|=5.33$ Å ($/\sqrt{2} = 3.77$ Å), and $|\textbf{c}_{111}|=26.17$ Å ($/4\sqrt{3} = 3.78$ Å). Therefore, for this phase, we remark that straining $\textbf{a}_{111}$, $\textbf{b}_{111}$ to the same pseudocubic lattice parameter as we do leads to marginally different in-plane strains.

\begin{table}[b]
\caption{$\Delta E$ of bulk relaxed NdNiO$_3$ phases wrt bulk, non-magnetic $Pm\bar{3}m$.
}
\label{tab:Bulk_Energies}
\begin{ruledtabular}
\begin{tabular}{ccc}
\textrm{Space Group} & \textrm{Mag. Order}&
\textrm{$E-E_{R\bar{3}m}$ (meV/f.u.)}\\
\colrule
$P2/c$ & $T$-AFM & -277 \\
$P2/c$ & $S$-AFM & -276 \\
$Pbnm$ & FM & -264 \\
$P\bar{1}$ & $T$-AFM & -259 \\
$P\bar{1}$ & $S$-AFM & -258 \\
$R\bar{3}c$ & FM & -246 \\
\end{tabular}
\end{ruledtabular}
\end{table}

\subsection{Compressively [111]-strained NdNiO$_3$}

\begin{figure}
    \centering
    \includegraphics[width=1\linewidth]{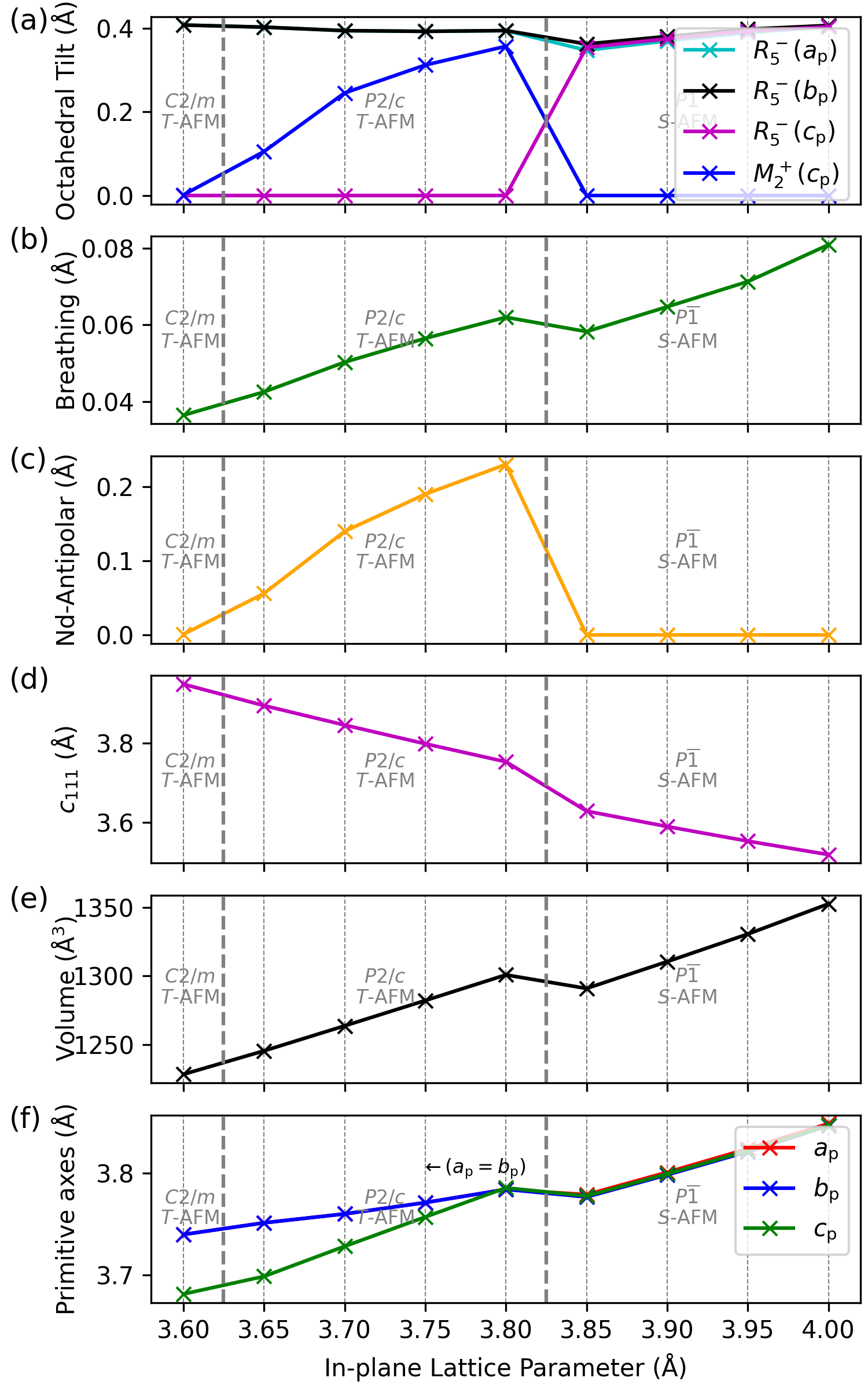}
    \caption{Structural properties as a function of [111]-oriented epitaxial strain: (a) octahedral tilt amplitude (where $a_{\text{p}}, b_{\text{p}}, c_{\text{p}}$ denote primitive axes normal to octahedral rotation), (b) breathing mode amplitude ($R_2^-$), (c) Nd-site anti-polar mode amplitude ($X^-_5$), (d) out-of-plane lattice parameter (in pseudocubic units), (e) cell volume and (f) lattice parameters along primitive axes (in pseudocubic units). 
    }
    \label{fig:Structural}
\end{figure}

The structural properties of NdNiO$_3$ as a function of [111]-strain are presented in Fig. \ref{fig:Structural}(a)---(f). The energies of all stable and metastable phases as a function of strain (that could be relaxed) are also included in Appendix B for reference. Starting with the compressive regime (3.60---3.75 Å), we observe the structure to remain in the $P2/c$ ($T$-AFM) space group, except for very high strains where the system transitions to a higher symmetry $C2/m$ ($T$-AFM) phase. As strain is applied, the in-phase $M_2^+$ tilt declines steadily, while the Nd-antipolar motion declines concomitantly, suggesting a similar coupling of antipolar motion to the $a^-a^-c^+$ tilt pattern as observed in $Pnma$ perovskites \cite{Orthorhombic_antipolar_coupling}. Interestingly however, the two $R_5^-$ tilts remain stable, even increasing slightly with strain. This could be explained by the competitive coupling between octahedral tilts, where the declining $M_2^+$ mode allows the other two tilts to increase in magnitude. The breathing mode also steadily declines throughout the compressive regime. In bulk, the appearance of the breathing mode is triggered by a critical amount of octahedral tilting \cite{Triggered_MIT}. It is plausible that the decline of the breathing mode seen here is driven by the decline in the $c^+$ tilt (thus reducing the net tilt), though as the two $a^-$ tilts increase slightly, it is hard to conclude that tilting is the only factor causing this. Despite the predictable increase of the out-of-plane parameter $|\textbf{c}_{111}|$, the three primitive axes decline with strain, particularly in the $\textbf{c}_{\text{p}}$ direction (likely a result of the vanishing tilt about this primitive axis). 
The variation in primitive parameters is driven primarily by [111]-strain directly, which applies an equal geometric constraint on each primitive direction (though this is complicated by the effects of tilt competition here).

\begin{figure}
    \centering
    \includegraphics[width=1\linewidth]{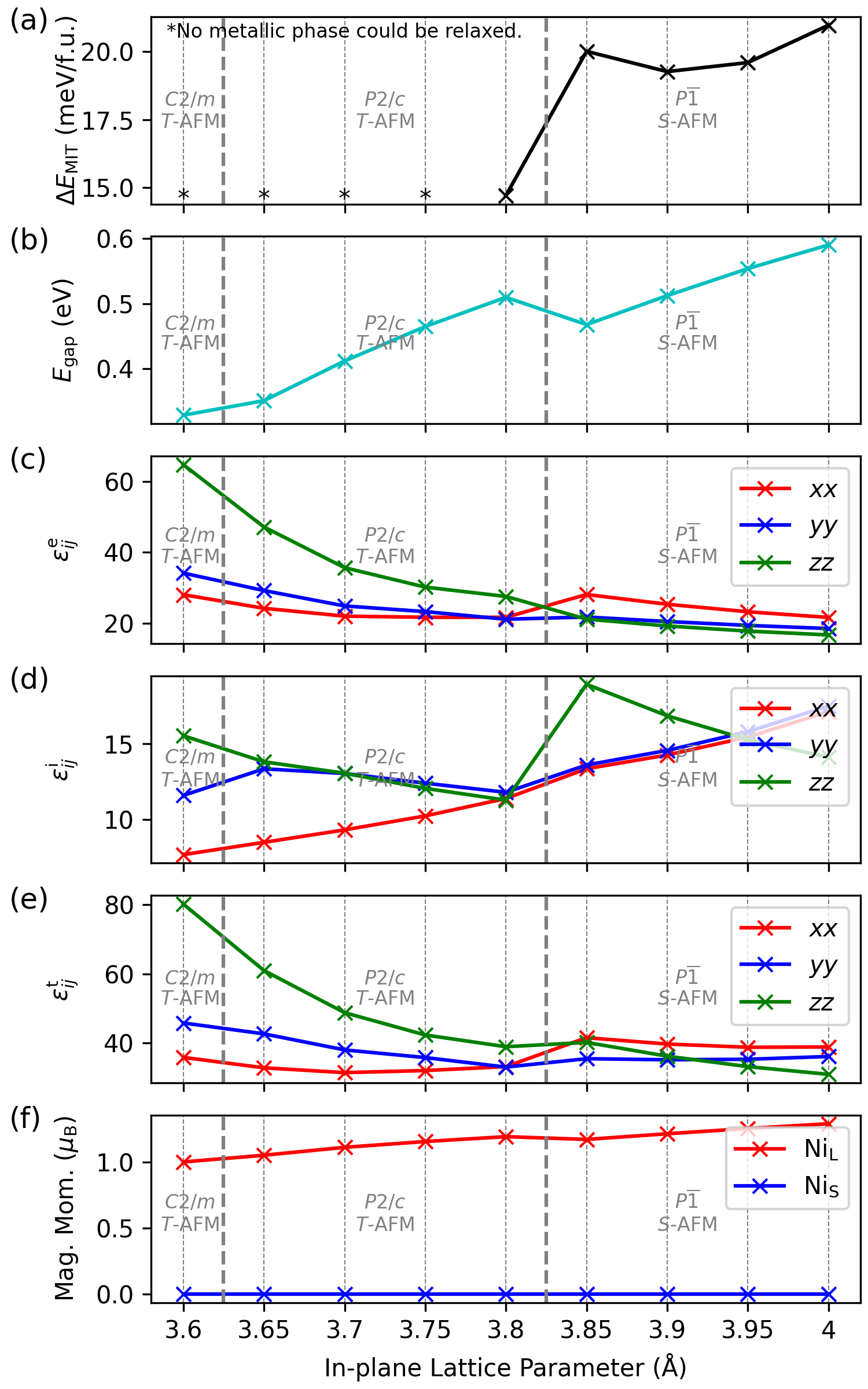}
    \caption{Electronic properties as a function of [111]-oriented epitaxial strain: (a) energy difference between lowest energy metallic phase and ground state, (b) electronic band gap, (c)---(e) electronic (superscript $e$), ionic (superscript $i$), and total (superscript $t$) static dielectric tensor components (where labels $\text{xx}, \text{yy},\text{zz}$ correspond to the components of the leading diagonal of the $3\times 3$ dielectric tensor $\epsilon_{ij}$, calculated along directions $\hat{\textbf{a}}_{111},\hat{\textbf{b}}_{111},\hat{\textbf{c}}_{111}$), (f) magnetic moments of large and small Ni sites.
    }
    \label{fig:Electronic}
\end{figure}

The electronic properties of NdNiO$_3$ as a function of [111]-strain are presented in Fig. \ref{fig:Electronic}(a)---(f). To estimate the MIT temperature, we plotted the energy differences $\Delta E_{\text{MIT}} = E_{\text{metal}} - E_{\text{insulator}}$ between the lowest energy metallic and insulating phases (Fig. \ref{fig:Electronic} (a)). We could not relax any metallic phases in the [111]-compressive regime, which is surprising, as the breathing mode is declining in magnitude. The width of the electronic band gap (\ref{fig:Electronic}(b)) is heavily coupled to the structural properties of NNO in bulk. The variation of the breathing mode amplitude directly modulates the band gap, while the tilt and Ni-O distance modulates the valence and conduction band dispersion (which moderately affects the band gap as well) \cite{Triggered_MIT}. In the compressive regime, considering the net tilt and breathing mode decline, it is no surprise therefore that the electronic band gap also declines. However the decline is relatively moderate, and thus quite far from closing and inducing an MIT. We also computed the dielectric properties (Fig. \ref{fig:Electronic} (c)---(e)) to check for polar instabilities, and found that in the [111]-compressive regime the dielectric component along the $\textbf{c}_{111}$ direction is amplified with increasing strain, while $\textbf{a}_{111}$ and $\textbf{b}_{111}$ components are also moderately amplified. This is primarily due to the electronic part of the response, which suggests that the system is in fact approaching the IMT. This appears to affect the $\textbf{c}_{111}$ direction more than the orthogonal directions--- likely a result of the elongation of this axis with compressive strain (Fig. \ref{fig:Structural}(d)). The ionic component is also slightly amplified along $\textbf{c}_{111}$, which indicates the system is also becoming mildly closer to a polar instability, undoubtedly also a result of increasing the aspect ratio of the cell (with respect to the [111]-basis). 

We also compare the effect of [111]-strain to that of [001]-strain, where we have reinvestigated the latter with the same parameters as our investigation into [111]-strain. Structural and electronic data for the effects of [001]-strain may be found in Appendix C. On the whole, in the compressive regime the effect of [111]-strain on the structure of NNO is not as extreme as under compressive [001]-strain, which completely suppresses the breathing mode at 3.70 Å. This can be explained by the difference in the geometric constraints on the primitive axes for each basis. [111]-strain affects the primitive axes equally, while [001]-strain splits the in-plane and the out-of-plane axes. The latter has a more substantial net effect on the tilts, particularly at high strains, where the IMT is induced. However, for medium strains, the net tilt does not change substantially, while breathing continually declines, suggesting that tilting is not the only cause of this. The electronic properties of the [111]-compressive regime are also comparable to those of the [001]-compressive regime. Right before the IMT, the electronic dielectric components diverge significantly, to a similar magnitude to the [111]-regime at our maximal compressive strain. This indicates that the [111]-strained system may actually be far closer to an IMT at 3.70 Å than suggested by the structural properties. The out-of-plane ionic component also increases under compressive [001]-strain, which is also likely driven by cell tetragonality. This is even despite the out-of-plane tilt increasing--- if this could be suppressed, it is likely that the polar mode could become unstable. However, as with the [111]-basis, reducing the tilts may quickly induce the IMT instead. Furthermore, octahedral tilt matching (particularly out-of-plane) to a substrate in the [001]-basis typically has a much weaker effect (due to connectivity of the octahedra). Therefore, the creation of a ferroelectric NNO phase in the [001]-basis is likely to be far more challenging than in the [111]-basis.

\subsection{Tensile [111]-strained NdNiO$_3$}

In the tensile regime (3.80---4.00 Å), initially (3.80 Å) the system remains in the $P2/c$ phase. However, beyond this point, we observe the structure to rapidly stabilise a triclinic $P\bar{1}$ ($S$-AFM) phase, where the $c^+$ tilt of the $P2/c$ phase switches to an anti-phase tilt of approximately the same magnitude as the other antiphase tilts ($a^-a^-a^-$). Simultaneously, the antipolar mode disappears, due to it being dependent on the $a^-a^-c^+$ tilt pattern. However, the breathing mode remains present, despite the small drop in amplitude. This suggests that the model of the triggered MIT in Ref. \cite{Triggered_MIT} may also apply to tilt patterns other than $a^-a^-c^+$, which is consistent with the results in Ref. \cite{Triggered_MIT} showing that both antiphase and in-phase tilts (separately) couple cooperatively to the breathing mode. 
The primitive parameters also all drop during this transition (as well as $c_{111}$ and the volume), which can be attributed to the change in tilt pattern and disappearance of the Nd antipolar mode. 
Therefore, though the breathing mode does not change drastically, we might guess that the $a^-a^-a^-$ tilt pattern is slightly less able (for the same tilt magnitudes) than the bulk $a^-a^-c^+$ pattern to trigger the breathing mode. Regardless, beyond this transition, we see the breathing mode and band gap steadily increase. The tilts, driven by strain, only increase minimally, which suggests that this is not the only factor affecting the breathing mode (which we explore in the next section).

The energy difference between metallic and insulating phases increases significantly in the [111]-tensile regime, even for low strains. It appears to tail off for medium tensile strain, but continues increasing beyond this. This matches nicely to experiment, where the MIT temperature has been observed to enhance significantly even at the low [111]-tensile strain enforced by an orthorhombic NdGaO$_3$ substrate \cite{111_NNO_expt}. The band gap steadily increases in the [111]-tensile regime, which can be attributed to the increase in breathing mode, as well as the slight increase in tilts. We note that based on above arguments regarding tilts, enforcing an $a^-a^-c^+$ pattern is likely to amplify the gap further, given the drop in magnitude when transitioning from $P2/c$ to $P\bar{1}$. The [111] dielectric properties remain stable, with no indication of a polar mode appearing with increased tensile strain. However, we notice a significant out-of-plane ionic dielectric component amplification at low strains, while the system remains in the $P\bar{1}$ phase. This suggests that phases with the $a^-a^-a^-$ tilt pattern are more cooperative with the polar mode than those with the $a^-a^-c^+$ pattern. The $a^-a^-a^-$ pattern has been shown to cooperate with ferroelectricity (forming the $R3c$ phase) in several bulk perovskites \cite{Coop_tilts_FE}. We continue the discussion of polar nickelate phases in a dedicated subsection below. One other interesting point to note is in regards the magnetic order. We find that in the tensile regime, the $S$-AFM order is lower in energy than the $T$-AFM order. Provided the description of these two magnetic orders as AFM ($S$-AFM) and FM ($T$-AFM) in-plane, this suggests that tensile [111]-strain favours AFM in-plane layers, while compressive [111]-strain favours FM (111) layers.

As with the compressive regime, there are many similarities between the [111]- and [001]-tensile regimes (Appendix C). In the latter, we see the breathing mode similarly increase. The tilts however remain more or less fixed in magnitude (which in itself is unusual for an [001]-strain phase diagram, as tensile [001]-strain typically allows for greater in-plane tilts at the cost of out-of-plane tilts). This suggests that as in the [111]-basis, there are separate driving forces behind breathing mode amplification besides the tilts in strained rare earth nickelates. A key difference between the two regimes is in regards the favouring of the IMT. At our maximal tensile strain (4.00 Å), we do not observe the system to undergo an IMT in the [001]-basis, though previous DFT studies report that higher strains can induce this \cite{Nickelate_001_Strain_DFT}. Our data for the energy difference between the lowest metallic and insulating phases support this theory, where $\Delta E_{\text{MIT}}$ rapidly decreases under [001]-tensile strain. This is despite the increase in the breathing mode, and much unlike the [111]-case, where the metal-insulator energy difference instead increases with tensile strain. [111]-tensile strain appears to be the only strain that shows no signs of the system undergoing an IMT, but rather quite the opposite. In regards the magnetic order, interestingly we also report similar findings to the above, where tensile [001]-strain favours $S$-AFM, and compressive strain favours $T$-AFM. A possible explanation for the driving force behind this alongside the strain itself is the tilts, where increasing the tilts appears to favour $S$-AFM, while decreasing them favours $T$-AFM.

\subsection{Investigation of a potential polar instability}

To see if we could induce a polar NNO phase, we investigated the ferroelectric instability of the metastable $P\bar{1}$ phase at 3.75 Å, the lowest compressive point where it could be relaxed (higher compressive strains relax the phase to a $C2/m$ ($a^-a^-c^0$) state). Here, we could not relax a polar mode, but found the out-of-plane ionic component of the dielectric permittivity to be 25.84, an increase from 3.85 Å (18.92). In an interfaced system, it may be possible via substrate tilt control to induce the $P\bar{1}$ phase deep into the compressive regime. At these points, the resulting system is likely to be closer to the polar instability, which could tentatively lead to an exotic ferroelectric, charge-ordered nickelate phase. On a separate note, substrate tilt matching could also reduce the NNO tilt magnitudes, making a polar phase more energetically favourable (if the tilts are reduced too much, this may prevent the triggering of the breathing mode, however). Experimentally, on [111]-LaAlO$_3$ (which applies only a minimal tensile strain to NNO, but enforces the $a^-a^-a^-$ tilt pattern), a polar (metallic) NNO phase has been observed--- which is attributed to tilt matching \cite{Polar_metals}. Our results indeed suggest that inducing this tilt pattern in NNO moves the system closer to a polar instability, while simultaneously suppressing the MIT. Our results also suggest that applying compressive [111]-strain could amplify this effect further. However, we emphasise that our calculations suggest that NNO is still likely to be quite far from being polar from the effect of [111]-strain alone. 

\subsection{Coupling of the breathing mode to strain}

\begin{figure}
    \centering
    \includegraphics[width=1\linewidth]{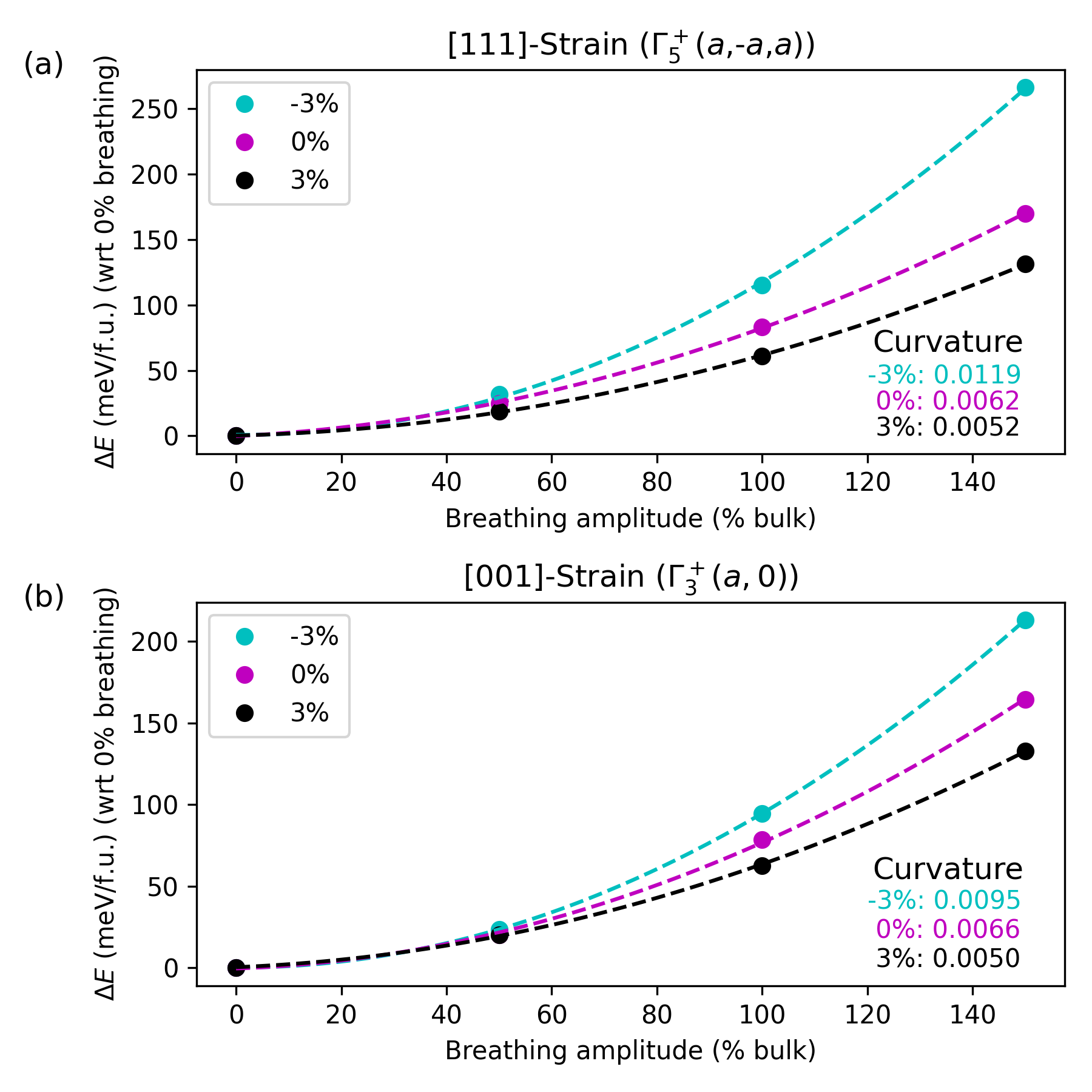}
    \caption{Energy landscape of the breathing mode, and associated energy curvature, as a function of (a) [111]-strain and (b) [001]-strain.
    }
    \label{fig:Breathing_Strain}
\end{figure}

Our results suggests that in both the [111]-compressive and tensile regimes (as well in the [001]-basis), there exists an alternative driving force for the appearance of the breathing mode, besides octahedral tilt. To clarify this, we investigated the mode coupling between the breathing mode and strain itself. For several strains, both [111] and [001], we measured the energy of the system when varying the breathing mode amplitude, from 0\% to 150\% of the bulk value. Our findings are shown in Fig. \ref{fig:Breathing_Strain}. For each curve, we measure the energy curvature. A smaller curvature indicates that the breathing mode is more energetically favourable, while a larger curvature indicates the reverse. Looking first at the effect of [111]-strain (Fig. \ref{fig:Breathing_Strain}(a)), we observe the energy curvature to increase with compressive strain, and decrease with tensile strain. This matches exactly what is seen in our strain phase diagram, where the breathing mode decreases and increases in magnitude with compressive and tensile strain respectively, but also shows that this variation is not only driven by tilt (which is also driven by strain), but also strain alone. The result for [001]-strain is very similar, where the curvature also increases and decreases in compressive and tensile strain respectively. This matches our [001]-strain phase diagram.

Interestingly however, though the magnitude of the curvature is greater under compressive [111]-strain than under compressive [001]-strain, in our strain phase diagrams, we only see an MIT in the [001]-compressive regime. This suggests that the IMT in the [001]-basis is primarily driven by the coupling between the strain and the tilts (which in-turn affects the breathing mode due to the tilt-breathing coupling) rather than the coupling of the strain and the breathing mode. Though we see a low [001]-tensile strain-breathing curvature, which suggests the stability of an insulating phase, in our strain phase diagram we find that $\Delta E_{\text{MIT}}$ declines with [001]-tensile strain. This is likely to be driven separately by the strong decrease in out-of-plane bond lengths (in contrast, Ni-Ni bond lengths all increase under [111]-tensile strain), which increases the band dispersion.

\section{\label{sec:level1}Summary}

Applying [111]-strain to rare-earth nickelates opens the door to several new phases and properties. We observe a new $C2/m$ phase induced under high compressive strain, and a new $P\bar{1}$ phase induced under tensile strain. [111]-tensile strain also steadily amplifies the band gap, and appears to increase the MIT temperature. We highlight that charge ordering in the rare-earth nickelates is not restricted to the $a^-a^-c^+$ tilt pattern (even without substrate tilt matching)--- it can also be triggered in alternative tilt patterns such as $a^-a^-a^-$, which suggests that charge ordering could be possible in other rhombohedral perovskites (such as LaNiO$_3$). Given the link between this tilt pattern and ferroelectricity, we also checked for the polar instability in our structures, though found that they remain far from turning polar under [111]-strain alone (though the phases with an $a^-a^-a^-$ tilt pattern had a higher dielectric susceptibility than those with an $a^-a^-c^+$ tilt pattern). Comparing the effect of [111]-strain to that of [001]-strain, we find that the rare-earth nickelates are generally less sensitive to the former, which matches recent studies investigating the effect of [111]-strain on other orthorhombic perovskites. We also clarify further the coupling between strain and charge ordering. We find that the energy penalty of the breathing mode, in both [001]- and [111]-bases, increases under compressive strain. This, combined with an understanding of the coupling of the breathing mode to tilts, and subsequent coupling to the band gap, allows us to explain trends in the structure and MIT with [111]- and [001]-strain. We hope that these results encourage further experimental investigation into the epitaxially [111]-strained rare-earth nickelates.

\begin{acknowledgements}
J. Í.-G. acknowledges the financial support from the Luxembourg National Research Fund through grant C21/MS/15799044/FERRODYNAMICS. 
A. L. acknowledges EPSRC for part-funding of his studentship (EP/T518001/1). 
This work used the Hamilton HPC service at Durham University. 
We thank Pavlo Zubko and Evgenios Stylianidis for helpful discussions.
\end{acknowledgements}

\appendix
\counterwithin{figure}{section}
\counterwithin{table}{section}

\section{Properties of bulk NdNiO$_3$}

Table \ref{tab:NNO_bulk} provides some basic structural and electronic properties of the ground state (bulk, fully relaxed and unstrained) of NNO in the $P2/c$ space group with $T$-type AFM ordering. Mode magnitudes and lattice parameters (see Appendix A for details) were found to be comparable to low temperature synchrotron data, and the band gap and magnetic moments comparable to previous DFT studies. 

\begin{table}[h]
\caption{Structural and electronic properties of bulk NNO ($P2/c$ $T$-AFM)}
\label{tab:NNO_bulk}
\begin{ruledtabular}
\begin{tabular}{cc}
Quantity & Value \\ \hline
Structural Modes (Å): & \\
$R_5^-(a_\text{p})$ & 0.39 \\ 
$R_5^-(b_\text{p})$ & 0.39 \\ 
$M_2^+(c_\text{p})$ & 0.33 \\ 
$R_2^-$ & 0.06 \\ 
$X_5^-$ & 0.21 \\ \hline
Lattice Parameters (prim. basis)& \\
$a_{\text{p}}$ (Å) & 3.78 \\ 
$b_{\text{p}}$ (Å) & 3.77 \\ 
$c_{\text{p}}$ (Å) & 3.78 \\ \hline
Volume, $V$ (Å$^3$) & 1290.85 \\ \hline
Band gap, $E_g$ (eV) & 0.47 \\ \hline
Dielectric Tensor ([111]-basis) & \\
$\chi_{xx}^e$ & 21.62 \\
$\chi_{yy}^e$ & 22.33 \\
$\chi_{zz}^e$ & 28.57 \\
$\chi_{xx}^i$ & 10.84 \\
$\chi_{yy}^i$ & 12.15 \\
$\chi_{zz}^i$ & 11.68 \\
$\chi_{xx}^t$ & 32.46 \\
$\chi_{yy}^t$ & 34.48 \\
$\chi_{zz}^t$ & 40.25 \\ \hline
Magnetic Moments ($\mu_B$) & \\
Ni$_\text{L}$ & 1.18 \\
Ni$_\text{S}$ & 0 \\

\end{tabular}
\end{ruledtabular}
\end{table}

\section{Energy of phases in [111]-strained NdNiO$_3$}

\begin{figure}[!h]
    \centering
    \includegraphics[width=1\linewidth]{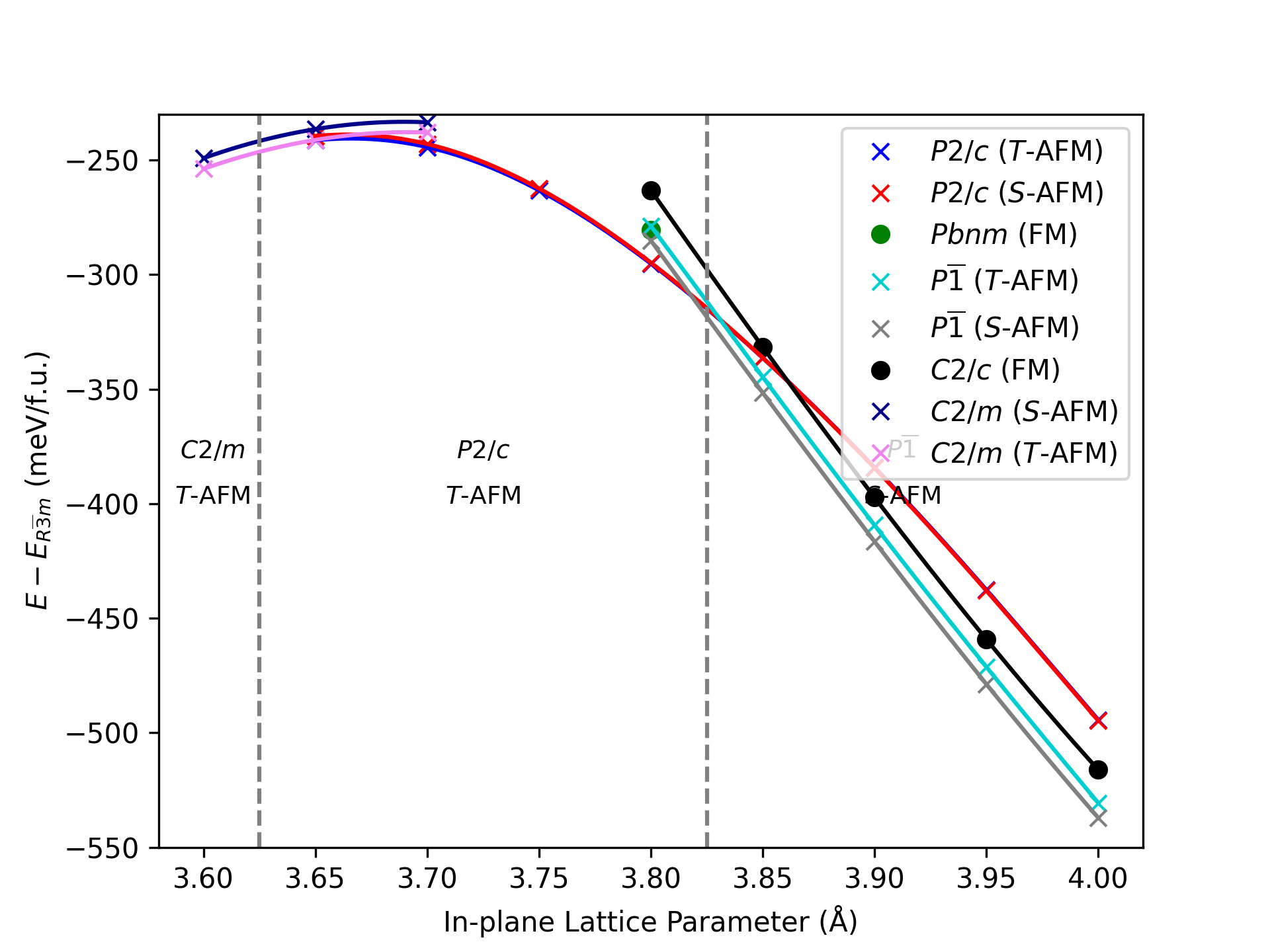}
    \caption{Energy of various NNO phases as a function of [111]-strain. Insulating and metallic phases are denoted by crosses and circles respectively. 
    }
    \label{fig:NNO_strain_phases}
\end{figure}

Fig. \ref{fig:NNO_strain_phases} provides the energy of various NNO phases and magnetic orders (that could be relaxed) with respect to the $R\bar{3}m$ phase ([111]-strained $Pm\bar{3}m$), as a function of [111]-strain. This data evidences the sequence of phase transitions, and critical strains, discussed in the main text. 

\newpage

\section{Structural and electronic properties of [001]-strained NdNiO$_3$}

\begin{figure}
    \centering
    \includegraphics[width=1\linewidth]{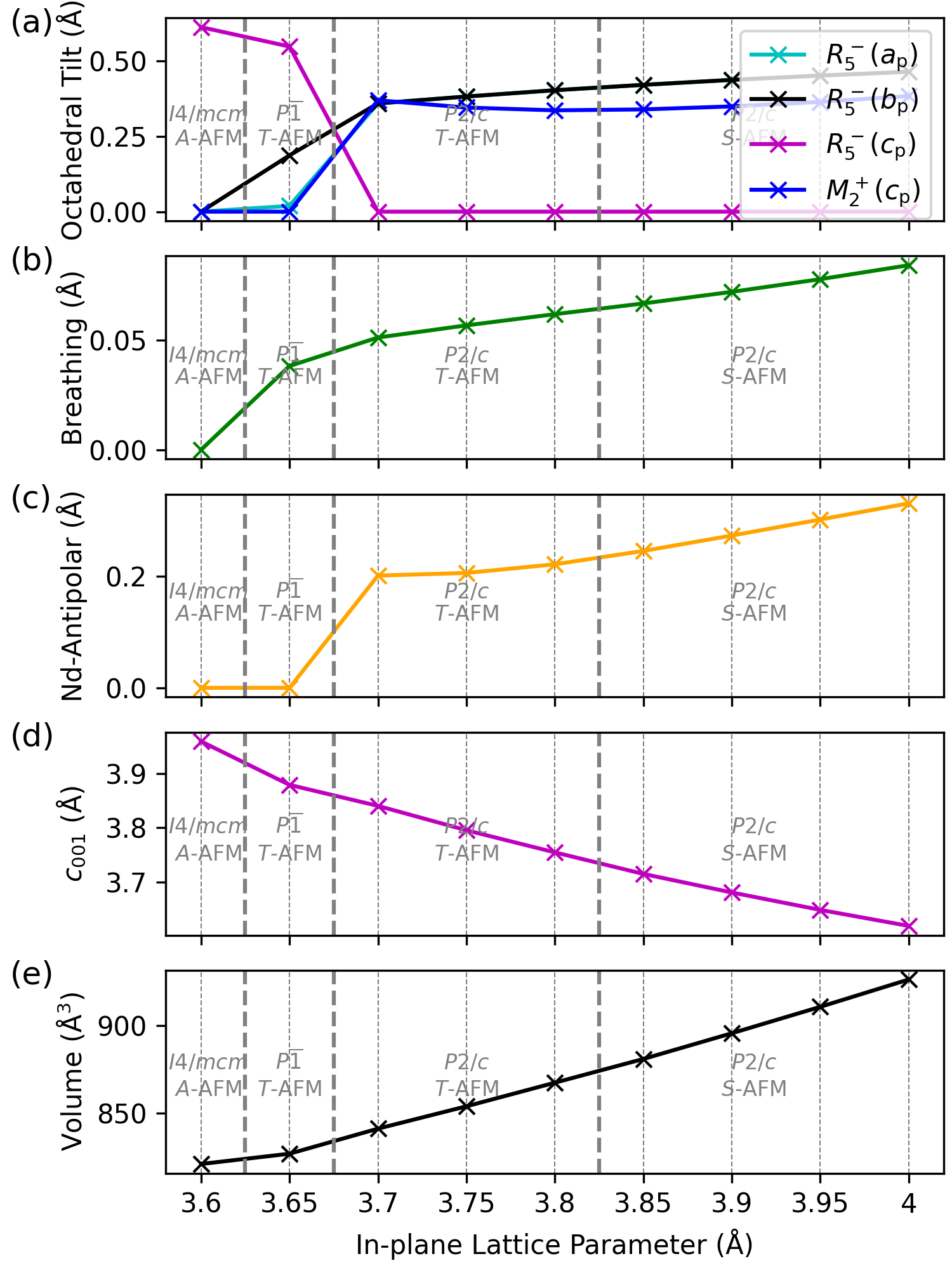}
    \caption{Structural Properties as a function of [001]-oriented epitaxial strain: (a) octahedral tilt amplitude (where $R_5^-$ denotes anti-phase tilting, $M_2^+$ denotes in-phase tilting, and $a_{\text{p}}, b_{\text{p}}, c_{\text{p}}$ denote primitive axes normal to octahedral rotation), (b) breathing mode amplitude ($R_2^-$), (c) Nd-site anti-polar mode amplitude ($X^-_5$), (d) out-of-plane lattice parameter (in pseudocubic units), (e) cell volume.
    }
    \label{fig:001_Structural}
\end{figure}

\begin{figure}
    \centering
    \includegraphics[width=1\linewidth]{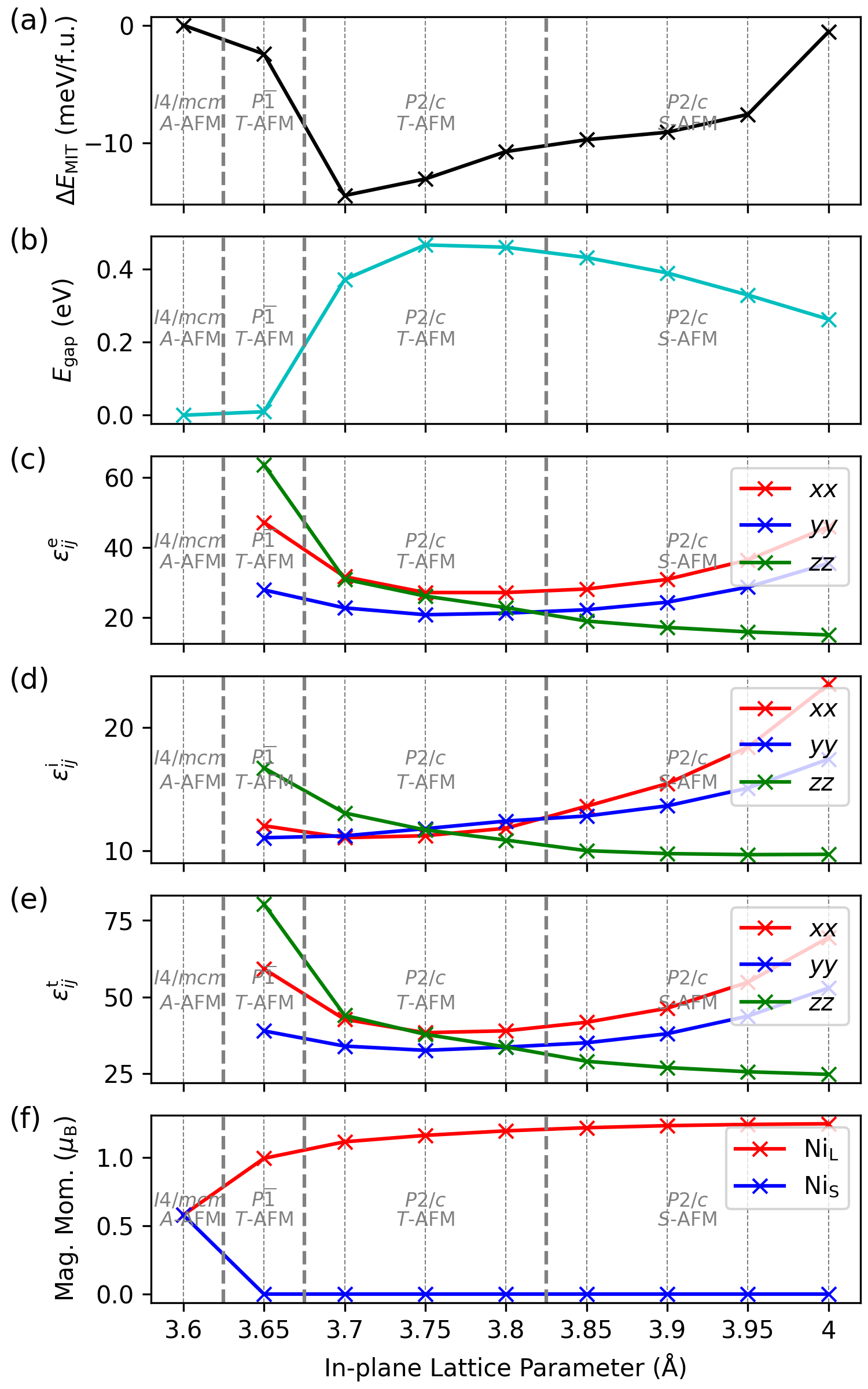}
    \caption{Electronic properties as a function of [001]-oriented epitaxial strain: (a) energy difference between lowest energy metallic phase and ground state, (b) electronic band gap, (c) electronic (superscript $e$), (d) ionic (superscript $i$), and (e) total (superscript $t$) static dielectric tensor components (where labels xx, yy, zz correspond to the components of the leading diagonal of the $3\times 3$ dielectric tensor $\epsilon_{ij}$, calculated along directions $\hat{\textbf{a}}_{001},\hat{\textbf{b}}_{001},\hat{\textbf{c}}_{001}$), (f) magnetic moments of large and small Ni sites.
    }
    \label{fig:001_Electronic}
\end{figure}

We reinvestigated the effect of [001]-oriented epitaxial strain on NNO, with results shown in figs \ref{fig:001_Structural} and \ref{fig:001_Electronic}. Here we note a few differences from previous DFT studies, with new phases appearing under compressive strain. For this study, we used the 80 atom Nickelate unit cell specified in Ref. \cite{Complete_Phase_Diagram_Nickelates}, and applied strain along the in-plane axes using the same method as in the [111] results. The plane-wave energy cut-off remained set to 900 eV, and the K-point grid was fixed to a $3\times 5 \times 2$ $\Gamma$-centered mesh. We also note that $A$-, $C$-, and $G$-type AFM orders \cite{ACG_AFM} were tested here within metallic phases, where we find $A$-AFM to be the ground state magnetic order for the metallic $I4/mcm$ phase at high compressive strain.

\bibliography{apssamp}

\end{document}